\documentclass[reprint,aps,prl,superscriptaddress,floatfix,showpacs]{revtex4-1}
\usepackage{times}
\usepackage{bm}
\usepackage{graphicx}
\usepackage{color}

\DeclareMathAlphabet{\mathscr}{OT1}{pzc}{m}{it}

\newcommand{\dens}[1]{{\rm [#1]}}

\newcommand{\tHe}{\mbox{$^3$He}}

\newcommand{\pinf}{P_{\infty}}

\newcommand{\be}{\begin{eqnarray}}
\newcommand{\ee}[1]{\label{#1}\end{eqnarray}}

\begin{document}

%\setpagewiselinenumbers
%\modulolinenumbers[5]
%\linenumbers

 \title{Breakdown of Angular Momentum Selection Rules in High Pressure Optical Pumping Experiments}

\def\Wisc{Department of Physics, University of Wisconsin-Madison, Madison, WI 53706, USA}
\def\Julich{Juelich Centre for Neutron Science, Garching 85747, Germany}
\author{B. Lancor}\affiliation{\Wisc}
\author{E. Babcock}\affiliation{\Julich}
\author{R. Wyllie}\affiliation{\Wisc}
\author{T. G. Walker}\affiliation{\Wisc}

\date{\today}

\begin{abstract}We present measurements, using two complementary methods,  of the breakdown of atomic angular momentum selection rules in He-broadened Rb vapor.  Atomic dark states are rendered weakly absorbing due to fine-structure mixing during Rb-He collisions. The effect substantially increases the photon demand for optical pumping of dense vapors.  
\end{abstract}

\maketitle

Optical pumping  \cite{Happer72,*HJW} of alkali-metal atoms at high temperatures and high buffer gas pressures is a powerful technique for precision spectroscopies (clocks, magnetometers, masers)  \cite{Jau04,*Kominis03,*Glenday08}, and for production of hyperpolarized noble gas nuclei by spin-exchange collisions   \cite{WalkerRMP}.  Hyperpolarized noble-gas nuclei are extensively used as neutron spin-filters  \cite{Gentile05,*Babcock09b}, targets for electron scattering  \cite{Singh09,*Slifer08}, for magnetic resonance imaging  \cite{Holmes08},  nuclear magnetic resonance  \cite{Zhou09}, and for fundamental studies  \cite{Morgan08,*Vasilakis09}.
For spin-exchange optical pumping, the small cross-sections for spin exchange  \cite{WalkerRMP}  are compensated for by operating at high alkali densities, producing optical depths of $D\sim 100$.  Such  extreme opacities are managed by pumping the atoms into ``dark'' Zeeman levels that do not absorb the pumping light.  Ideally, the vapor would  become completely transparent to the pumping light, except for a small unpolarized  layer near the cell walls.  Including collisional spin-relaxation losses would lead to a modest linear attenuation of the pumping light as it propagates through the cell  \cite{WalkerRMP}.  

Under these extreme conditions the quality of the dark state becomes of great importance.  In order to maintain high population of the dark state throughout the cell, the rate $R$ at which unpolarized atoms at the entrance to the cell absorb pumping light must be $\sim D\Gamma$, where $\Gamma$ is the spin-relaxation rate. Now suppose the dark state atoms absorb the light at a small rate $R_d$.  This directly increases the light absorption and also causes a small fraction of the atoms to populate strongly absorbing states.  Under these conditions, the  laser power required to optically pump the entire cell (the photon demand) increases by a factor \be\Upsilon=1+2R_d/\Gamma=1+2DR_d/R.\ee{a}
For typical $D=100$, a  dark state absorption rate of only 1/200 of the unpolarized absorption rate doubles the photon demand.

In this Letter we present the first observations of  dark-state absorption due to fine-structure mixing in alkali-metal He collisions.  We show that this light-induced spin-relaxation mechanism limits the attainable alkali polarization and substantially increases the photon demand for optical pumping.  While these collisions are well-studied in the context of line-broadening \cite{Allard82} and fine-structure population transfer \cite{Rotondaro98}, prior line-shape studies 
were insensitive to how the collisions alter the angular momentum selection rules for light absorption \cite{Romalis97}.

The dark state for optical pumping of alkali-metal atoms with light of photon spin projection $p=1$ tuned to the first excited $^2$P$_{1/2}$ level  is a fully nuclear ($I$) and electron ($S$) spin-polarized state with projection $m=S+I$.  It follows from angular momentum conservation and the lack of an excited Zeeman sublevel of projection $m'=S+I+1$ that excitation from the dark state is forbidden.   In contrast, the increased angular momentum of a $^2$P$_{3/2}$ level allows no such dark state when pumping to that excited level  \cite{Happer87}.

\begin{figure}
\includegraphics[width=3.0 in]{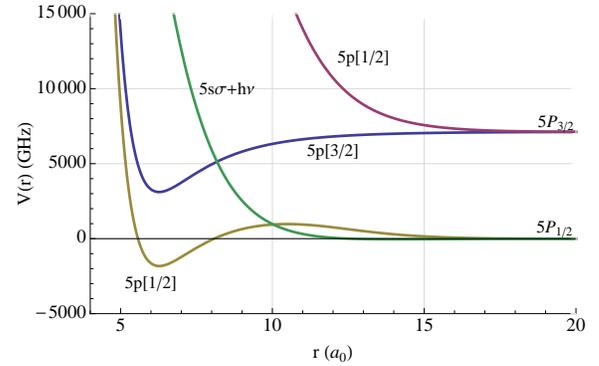}
\caption{Adiabatic energy curves for RbHe molecules  \cite{Pascale83} adapted to include fine-structure \protect \cite{Allard82}.  The projection  $|\Omega|$ of the total electronic angular momentum along the interatomic axis is given in brackets.  The curve crossings between the photon-dressed 5s$\sigma$ state and the two excited-state curves mean that photon absorption is allowed during a collision.  These absorption processes are not subject to the free-atom dipole selection rules, allowing normally angular momentum forbidden transitions to occur. }
\label{fig:pots}
\end{figure}

To the extent that collisions of excited alkali-metal atoms with the buffer-gas atoms do not mix states of different electronic angular momentum, the above arguments are unchanged in a buffer gas.  However, interactions with He   substantially mix the $^2$P$_{1/2}$ and $^2$P$_{3/2}$ states  even for Rb and Cs atoms with large fine-structure splitting  \cite{Allard82}.  This mixing appears in Fig.~\ref{fig:pots} as a divergence of the $|\Omega|=1/2$ adiabatic potentials, $\Omega$ being the projection of the electronic angular momentum along the interatomic axis.  In addition, the strong repulsion between $^2$S$_{1/2}$ atoms and He produces a transient absorption to the $|\Omega|=3/2$ state of purely $^2$P$_{3/2}$ character at an interatomic distance of 8a$_0$.  Both of these effects lead to substantial absorption of the pumping light from the dark state in the atmospheric pressure cells typically used for these experiments.  Thus the usual atomic angular momentum selection rules are violated in Rb-He collisions.

As described below, we have measured the dark-state absorption cross section for Rb in the presence of $^3$He gas of density $\dens{He}$ to be $\sigma_d=\dens{He}\times 1.21\pm0.12\times 10^{-17}$ cm$^2$/amg \cite{amg} for light near the $^2$S$_{1/2}\rightarrow ^2$P$_{1/2}$ resonance at a wavelength of 795 nm.   The consequences of this number can be understood by  calculating the photon demand for a detuned monochromatic pump laser.  With a pressure-broadened atomic lineshape of width $B'\dens{He}$, the photon demand doubles for light detunings of \be
\Delta=\sqrt{r_e f c B'\over 4 D \sigma_d}=104\mbox{ GHz}\sqrt{100\over D},
\ee{b}
where we have assumed $\Delta\gg B'\dens{He}$, and $B'=18.7$ GHz/amg \cite{Romalis97}.  The oscillator strength is $f$ and $r_e$ is the classical electron radius. Thus  light sources with widths exceeding 100 GHz will be very inefficient, at least  partially explaining why the narrowing of commercial high power diode lasers has produced substantial improvements in spin-exchange optical pumping \cite{Chen07}.  When light of broader bandwidth is used,  the on-resonant part of the pumping spectrum is rapidly depleted near the cell entrance, causing  the pumping spectrum to become increasingly off-resonant and therefore inefficient as the light propagates through the cell.

\begin{figure}[htb]
\includegraphics[width=3.0in]{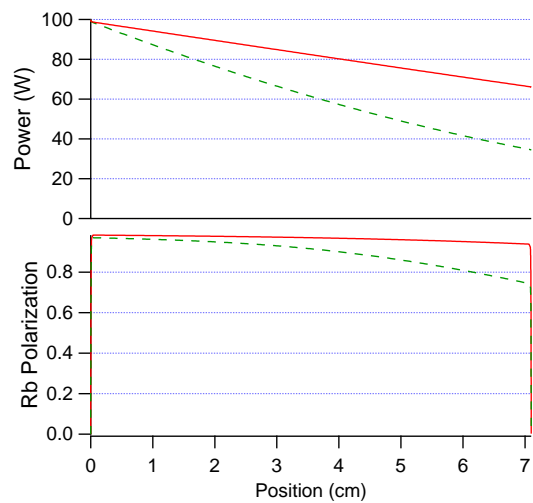}
\caption{Light propagation results with no (red, solid) and measured (green, dashed) dark-state absorption.  The upper graph shows the power as a function of position.  The dark state absorption produces a faster attenuation of the light.  The lower graph shows the corresponding Rb polarizations.  The dark state absorption reduces the Rb polarization at the cell entrance, with further reductions as the spectral profile of the light becomes  increasingly off-resonant deeper inside the cell. }
\label{fig:broad}
\end{figure}

Fig.~\ref{fig:broad} illustrates the dark state absorption effect, simulating the alkali polarization and laser power as a function of position in the cell, both with and without the dark state absorption.
We assume 100 W of  pumping light with a 1000 GHz FWHM spectral profile,  a 10 cm diameter, 7 cm long cell with 8 amg of \tHe\ and 0.066 amg of N$_2$, and \dens{Rb}=$4\times 10^{14}$ cm$^{-3}$.  
Under these conditions we estimate a spin-relaxation rate of 630/s  \cite{Chen07}.  Without dark-state absorption the light is attenuated only due to ground-state spin-relaxation and only 35 W would be dissipated in the cell,  maintaining  a very high Rb polarization. When the dark-state absorption is taken into account, several changes occur.  The power dissipation per unit length is substantially increased, as seen in Fig.~\ref{fig:broad}, even at the entrance to the cell before the spectral hole is burned.  The average power dissipation is much greater than in the ideal case, now 65 W.  The polarization drop is now quite substantial, reducing to 75\% at the back of the cell.  This is due to two effects:  1) the pumping rate is lower due to the greater power dissipation  and the production of a complete hole in the spectral profile, and 2) the remaining light is in the spectral region with low efficiency, further reducing the maximum attainable polarization.

We deduced the dark-state absorption cross sections using the change in transmission of circularly polarized light through an optically pumped vapor as the atomic spin-pol\-arization is reversed.  This is effectively a measurement of the circular dichroism of the vapor. Combining this with previous measurements of the absorption cross sections for unpolarized atoms  \cite{Romalis97} allows us to avoid a measurement of the Rb vapor pressure.

\begin{figure}[ht]
\includegraphics[width=3.0 in]{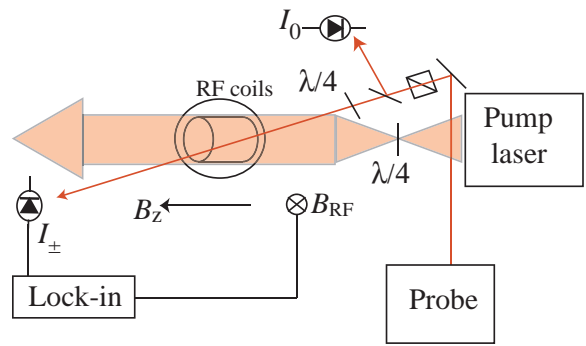}
\caption{Apparatus for measuring the circular dichroism of Rb-He vapor.Ê The pump laser, propagating parallel to a magnetic field $B_z$, spin-polarizes Rb atoms either parallel or antiparallel to the field by optical pumping.   The fractional transmission of a weak, circularly polarized, tunable probe laser is determined by the ratio of photodiode voltages before and after  the cell.Ê The circular dichroism is then determined from the transmissions for both directions of Rb polarization.Ê The absolute Rb polarization is determined by driving RF resonances with field $B_{\rm RF}$ and measuring the resulting modulated Faraday rotation of the probe laser. }
\label{fig:exp}
\end{figure}

As shown in ÊFig.~\ref{fig:exp}, a Rb vapor cell was optically pumped by a circularly polarized frequency narrowed diode array bar \cite{Babcock05}. ÊA magnetic field of 50 G was applied in the pump propagation direction. ÊA tunable probe external-cavity diode laser was attenuated to P$< 50$ $\mu$W, sent through a mechanical chopper operating at 485 Hz, and linearly polarized with a polarizing beam splitter cube. ÊA portion was split off by a non-polarizing beam splitter plate to provide a measure of the incident intensity, and the remainder was circularly polarized with a quarter wave plate. The probe beam propagated through the cell at an angle $\theta=17.6^\circ$ with respect to the magnetic field. ÊThe incident and transmitted intensities were sent to lock-in amplifiers referenced to the chopper frequency. ÊTo change the direction of the atomic spin polarization, the pump ${\lambda/4}$ plate was rotated 90$^\circ$. ÊTo obtain the relation between the incident and transmitted intensities Êin the absence of Rb (thus accounting for loss in the windows of the oven and cell), a measurement was taken at room temperature. Ê

Two natural abundance Rb cells were used in this experiment. ÊThe low pressure cell is a closed 4.9 cm long cylinder, with 0.80 amg $^3$He and 0.07 amg  of N$_2$.ÊThe high pressure cell  is a blown GE180 sphere of diameter 3.5 cm, Êfilled with 3.27 amg of $^3$He and 0.13 amg N$_2$. ÊThe cell being studied was placed in temperature-controlled  flowing hot air oven. ÊÊTemperatures ranging from $\sim$ 60 $^\circ$C to $\sim$ 180 $^\circ$C, corresponding to [Rb]= $1-200\times 10^{12}$ cm$^{-3}$, were used to produce appropriate optical thickness for transmission measurements at a range of frequencies. 

The Rb spin polarization Êwas measured with transverse electron paramagnetic resonance (EPR) spectroscopy \cite{Baranga98,Appelt98}. ÊA 26.4 MHz RF field was applied perpendicular to the holding field by driving a pair of 9 cm diameter coils, separated by 7 cm, with a Êsynthesized function generator. As the holding field was swept through the EPR resonances, the resulting polarization modulation of the probe beam  was measured by a polarizer and a fast photodetector, which was  demodulated to 100 KHz by mixing it with a 26.5 MHz signal from a second synthesized function generator. ÊThis 100 KHz signal was then sent to a lock-in amplifier referenced to a signal generated by mixing the outputs of the two function generators. ÊThe phase of the lock-in was chosen to produce Lorentzian EPR spectra. ÊThe spin-polarization was deduced using the area ratio method \cite{Baranga98}. Ê

\begin{figure}[tbh]
\includegraphics[width=3.5 in]{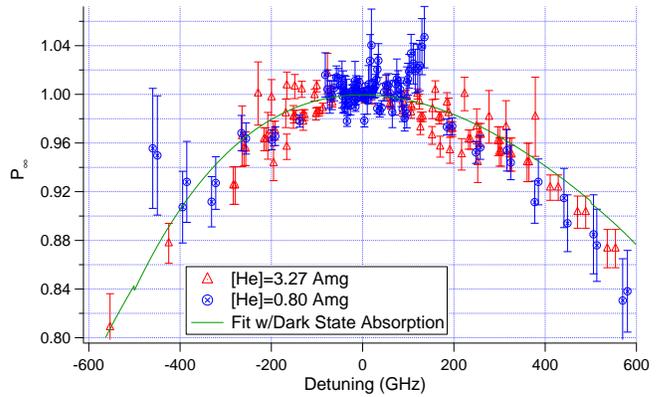}
\caption{Normalized circular dichroism results near the 5S$_{1/2}$-5P$_{1/2}$ resonance line of Rb. ÊThe agreement between cells of different He pressure verifies that the effect originates from absorption in RbHe collisions. ÊThe solid line shows the frequency dependence of the dichroism due to the dark state absorption. }
\label{fig:res2}
\end{figure}

The basic parameters observed by the experiment are the transmissions $I_\pm$ for  circularly polarized light propagating at angle $\theta$ to the spin polarization $\pm P$. ÊThe absorption cross section for the probe light is
\be
\sigma(\pm)=\sigma_0(1\mp P_\infty  P \cos\theta)
\ee{sigtheta}
where $P_\infty$, the normalized circular dichroism of the vapor,  would be 1  in the absence of the dark state absorption. 
We Êextract $\pinf$ from the transmitted intensities $I_\pm=I_o\exp(-\dens{Rb}\sigma(\pm)l)$ by finding 
\be
\frac{-\ln\left({I_{-}/I_o}\right)+\ln\left({I_{+}/I_o}\right)}{-\ln\left({I_{-}/I_o}\right)-\ln\left({I_{+}/I_o}\right)}=P P_\infty \cos\theta.
\ee{asymmetry}
Note that in forming this ratio the optical thickness $\dens{Rb}l$ and instrumental gains cancel.  We then use the previously measured pressure-broadened lineshape  \cite{Romalis97} $\sigma_0$ to obtain
\be
\sigma_d=\sigma_0(1-\pinf)
\ee{sigd}

The measured  $\pinf$ near resonance, from transmission data, for both cells is shown in ÊFig.~\ref{fig:res2}. ÊÊA very important result from Fig.~\ref{fig:res2} is the Êagreement between the two cells, despite their very different He pressures. ÊAt detunings outside the atomic linewidth of 15-60 GHz  \cite{Romalis97}, $\sigma_0$ is proportional to the He pressure. ÊThus, only if $\sigma_d$ is also proportional to the He pressure will the dichroism be pressure independent. ÊThe agreement between the two cells at different pressures thus confirms the source of the impure dichroism as being Rb-He collisions.  Other systematic checks of hyperfine effects, off-resonant pumping, and N$_2$ contributions, will be presented in a subsequent publication.  The solid line in Fig.~\ref{fig:exp} corresponds to $\sigma_d/\dens{He}=1.21\pm0.12$ cm$^2$/amg.  This value was used in the simulation of Fig.~\ref{fig:broad} which shows the dramatic effect of the breakdown of atomic selection rules on optical pumping of optically thick vapors.

\begin{figure}[htb]
\includegraphics[width=3.0in]{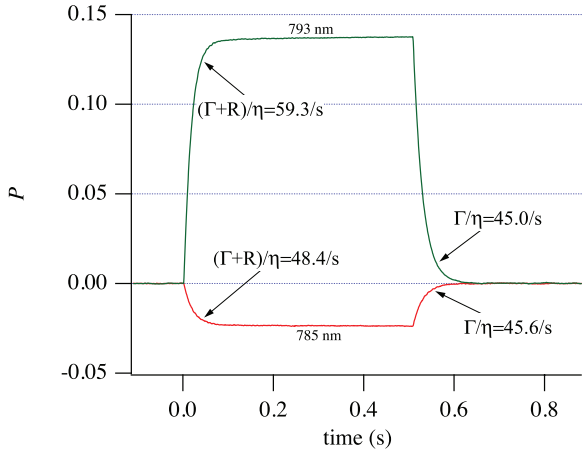}
\caption{Pumping/decay transients used to measure $\pinf$. Pumping light is turned on at $t=0$, and the polarization builds up to the steady-state value of Eq.~\ref{eqpol}.  The exponential build-up and decay constants allow the pumping rate $R$ and the relaxation rate $\Gamma$ to be measured.  Note that the sign of the polarization is opposite for 785 nm pumping as compared to 793 nm pumping, due to the dichroism of the vapor being negative for the former case. }
\label{fig:EarlTransient}
\end{figure}

As a second, quite different, method for measuring $\pinf$ at frequencies off the D1 resonance, we used a 30 W frequency narrowed  external cavity laser  \cite{Babcock05} to optically pump the atoms at different frequencies $\nu$.  The equilibrium polarization attained using optical pumping with light of normalized circular dichroism $P_\infty$ is
\be
P(\nu)=\pinf(\nu){R(\nu)\over R(\nu)+\Gamma}
\ee{eqpol}
Thus by measuring  $P(\nu)$, pumping rate $R(\nu)$, and spin-relaxation rate $\Gamma$, we deduce $\pinf$.

We chopped the pumping laser with a mechanical shutter and measured the spin-polarization as a function of time, as illustrated in Fig.~\ref{fig:EarlTransient}, using Faraday rotation.  
For small polarizations, the rising transient  builds up polarization to the steady state value (\ref{eqpol}) at a rate  $(R(\nu)+\Gamma)/\eta$, and the falling transient decays at the rate $\Gamma/\eta$, where the slowing-down factor $\eta=10.8$ for natural abundance Rb takes into account the spin inertia due to the alkali-metal nuclei at low polarizations \cite{Appelt98}.  The Faraday rotation was calibrated by EPR spectroscopy. 

\begin{figure}
\includegraphics[width=3.5 in]{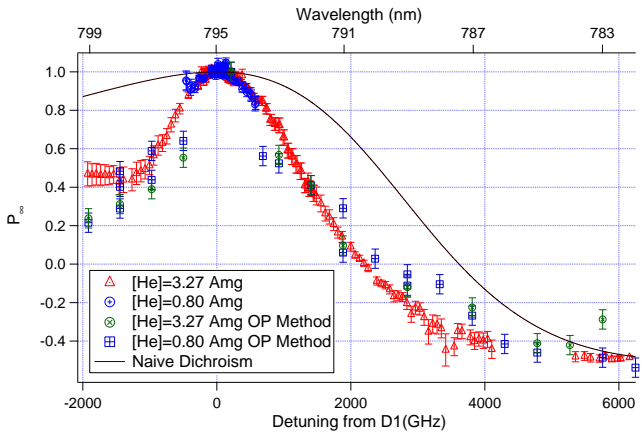}
\caption{Measured normalized circular dichroism of RbHe molecules in the region between the Êfirst resonance lines of Rb. ÊThe solid curve shows the expected dichroism from the very naive assumption of purely Lorentzian broadened lines. }
\label{fig:wide2}
\end{figure}

The deduced values of $\pinf$ using this optical pumping (OP) method are shown in Fig.~\ref{fig:wide2} and agree with the results of the direct dichroism measurements. 
With the OP method,  the zero crossing of the dichroism  is particularly dramatic as the signal of Fig.~\ref{fig:EarlTransient}
reverses sign near 790 nm. 
 We note that a naive pressure broadening model that neglects the angular momentum altering properties of RbHe collisions would predict the zero crossing to occur at 787.5 nm.

In summary, we used two distinct methods to demonstrate the breakdown of atomic angular momentum selection rules due to buffer gas collisions.  This breakdown  compromises the atomic dark state and has dramatic effects on the performance of optical pumping experiments with dense vapors.  To mitigate this effect, it is clearly important to perform the optical pumping with narrowband light tuned close to the atomic resonance.

\begin{acknowledgments}
We benefited from discussions with T. Gentile. This work was supported by the Department of Energy, Basic Energy Sciences.
\end{acknowledgments}

\bibliography{/Users/Thad_Walker/Research/thadbibtex/spinexchange}
%\bibliography{dichroism}

\end{document}